\documentclass{LT23auth}
\usepackage{graphicx}

\begin{document}

\begin{frontmatter}

\title{Strong evidence for the three-dimensional Fermi liquid behaviour of
quasiparticles in high-$T_{\rm C}$ cupurates}

\author{Setsuo Misawa\thanksref{thank1}}

\address{Institute of Quantum Science, Nihon University, Kanda-Surugadai, Tokyo 101-8308, Japan}

\thanks[thank1]{Corresponding author. Tel and Fax: +81-1-3259-0648\newline 
 E-mail: misawa@phys.cst.nihon-u.ac.jp}

\begin{abstract}
It is generally believed that behaviours of quasiparticles (holes) in high-
$T_{\rm C}$ cupurates should be attributed to the two-dimensional(2D) electronic states in the CuO$_{2}$ planes. The various anomalies of the transport coefficients for temperatures above $T_{\rm C}$ are long-standing insoluble puzzles and cause serious controversy. Here we reanalyse the published experimental date of LSCO cupurates. We find that the normal-state susceptibility, resistivity, Hall coefficient etc vary precisely as $T^{2}\ln T$ as a function of temperature $T$ in agreement with the prediction of the Fermi liquid model. The quasiparticles are shown to definitely behave as a 3D Fermi liquid. Various attempts to describe the system in terms of non-Fermi liquids,e.g. the RVB state, seem to be erroneous.
\end{abstract}

\end{frontmatter}

High-$T_{\rm C}$ cupurates are characterized by their anomalous properties 
in the normal state. The normal state resistivity often varies almost 
linearly with temperature $T$. 
To clarify the true nature of the normal state is a clue to understanding 
the superconductivity of this system. Many authors claim that the anomalies 
should be attributed to non-Fermi liquids. In contrast to this, here we 
present strong and conclusive evidence to show that quasiparticles (holes) 
of the system form a three-dimensional(3D) Fermi liquid, and hence the 
anomalous properties can be precisely described by the Fermi liquid model 
proposed by the present author in 1970 \cite{AC91}. Although the system is highly anisotropic, as will be seen below, the 3D nature of the quasiparticles is of critical importance. We have already shown that this
 model explains universally the physics of paramagnetic d-metals, 
 MnSi(FeSi), Laves phase compounds and heavy fermion compounds.

Because of the presence of the Fermi surface and of the interactions, 3D Fermi liquids produce various logarithmic (nonanalytic) terms; the quasiparticle energy contains, as a function of momentum $p$ with the Fermi 
momentum $p_{0}$, a $(p-p_{0})^{3}\ln |p-p_{0}|$ term which 
creates a $\varepsilon ^{2}\ln |\varepsilon|$ term in the density of states 
function, where $\varepsilon$ is the energy measured from the Fermi surface.
Thus the electrical conductivity due to impurities and/or imperfections is given by $\sigma _{\rm imp}=\sigma _{0} +\sigma _{1} T^{2} \ln (T/T_{1})$.
The total resistivity $\rho$ is a sum of the proper $\gamma T^{2}$
 resistivity arising from the Umklapp Coulomb scattering and the impurity resistivity $\rho _{\rm imp}={\sigma _{\rm imp}}^{-1}$; 
 $\rho (T)=\rho _{0} -\rho _{1} T^{2} \ln (T/T_{\rho}^{*})$, 
 where $\rho _{0}={\sigma _{0}}^{-1}$,
 $\rho _{1}=\sigma _{1}/{\sigma _{0}}^{2}$ and 
 $T_{\rho}^{*}=T_{1}\exp ({\sigma _{0}}^{2}\gamma /\sigma_{1})$.
 In contrast to the general belief, $\rho$ varies as 
 $T^{2}\ln (T/T_{\rho}^{*})$, where temperature $T_{\rho}^{*}$ 
(or $T_{1}$) has no 
 particular physical meaning and takes any value, for it arises from a sum of $T^{2}$ and $T^{2}\ln T$ terms.  
 Here it is important to note that 
 $T^{2}\ln (T/T_{\rho}^{*})$ varies almost linearly with $T$ for 
 $0.1\!$\hspace{0.3em}\raisebox{0.4ex}{$<$}
       \hspace{-1.1em}\raisebox{-.7ex}{$\sim$}\hspace{0.3em}$\!\!\!$
       $T/T_{\rho}^{*}\!$\hspace{0.3em}\raisebox{0.4ex}{$<$}
       \hspace{-1.1em}\raisebox{-.7ex}{$\sim$}\hspace{0.3em}$\!0.4$. 
By the same reasoning, the Hall 
 coefficient varies as $T^{2}\ln T$. Because of the short-wavelength density fluctuations the spin antisymmetric part of the Landau {\it f}-function contains a 
 $(p-p_{0})^{2}\ln |p-p_{0}|$ term which causes the $T^{2}\ln T$
 variation of the magnetic susceptibility 
 $\chi$; $\chi (T)=\chi (0)-bT^{2}\ln (T/T^{*});$ $\chi (T)$ should exhibit a maximum at temperature $T_{\rm max}=T^{*}/\sqrt{e}$.
 Here again $T_{\rm max}$ has no particular meaning; it does not represent the strength of antiferromagnetic correlations.
 
 To confirm the above view, we reanalyse the published experimental data
  of ${\rm La}_{2-x}{\rm Sr}_{x}{\rm Cu}{\rm O}_{4}$ 
  cupurates for a wide range of $x$. References \cite{AC92,AC93} are selected on 
  the basis of good samples and mutual consistency of the data. 
  Figure 1 shows that the experimental data of $d\rho /dT$ follow precisely 
  a $T \ln T$ law, and hence $  \rho -\rho _{0}\sim T^{2}\ln T$, up to about 
  $500{\rm K}$ above which higher order terms are needed. In fact, as shown 
  in Fig.2, the resistivity data can be precisely fitted by 
  $\rho(T)=\rho _{0}-\rho _{1}T^{2}\ln (T/T_{\rho}^{*})
  -\rho _{2}T^{4}\ln (T/T_{\rho}^{**})$ up to $1000{\rm K}$. The latest experimental data \cite{AC94} of the in-plane and out-of-plane resistivities for high quality single crystals also support this law. Similarly the Hall coefficient and Hall angle are shown to vary as $T^{2}\ln T$. 
  Figure 3 shows that the variation of $\chi (T)$ follows a $T^{2}\ln T$
   law; if the system were a 2D Fermi liquid, $\chi (T)$ would vary linearly with $T$ \cite{AC95}. Surprisingly this law is valid continuously up to $x=0$; 
   some parts of holes in ${\rm La}_{2}{\rm Cu}{\rm O}_{4}$
   behave as a 3D Fermi liquid.
   
   In conclusion, the normal state properties of high-$T_{\rm C}$ cupurates can be thoroughly described within the framework of the 3D Fermi liquid.
   The concept of non-Fermi liquids is not needed at all.
   \begin{figure}[btp]
\begin{center}\leavevmode
\includegraphics[width=\linewidth]{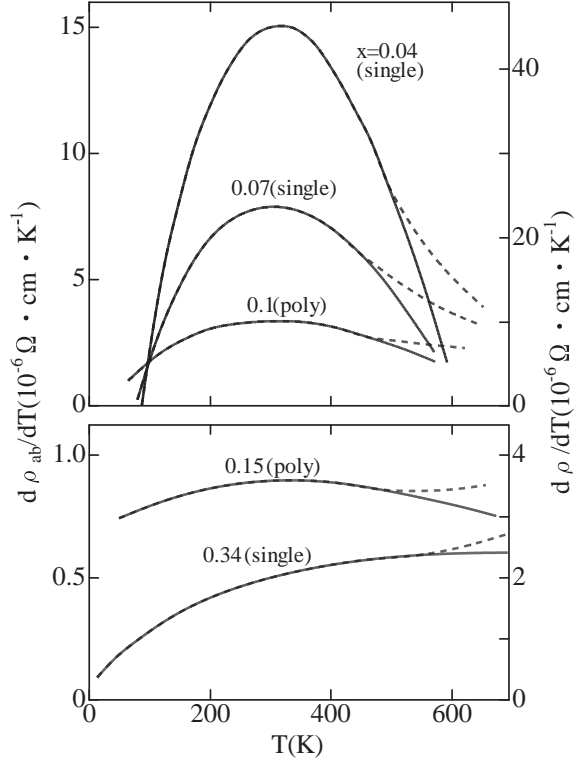}
\caption{ 
$d\rho /dT$ vs $T$ of single-crystal (left scale) 
        and polycrystalline (right scale)
        ${\rm La}_{2-x}{\rm Sr}_{x}{\rm Cu}{\rm O}_{4}$.
        The smoothed experimental date \cite{AC93} fit precisely theoretical $T\ln T$
        curves (solid lines) up to about $500{\rm K}$ above which the date 
        (dashed lines)  deviate from the solid lines.
}\label{figurename}\end{center}\end{figure}

\begin{figure}[btp]
\begin{center}\leavevmode
\includegraphics[width=\linewidth]{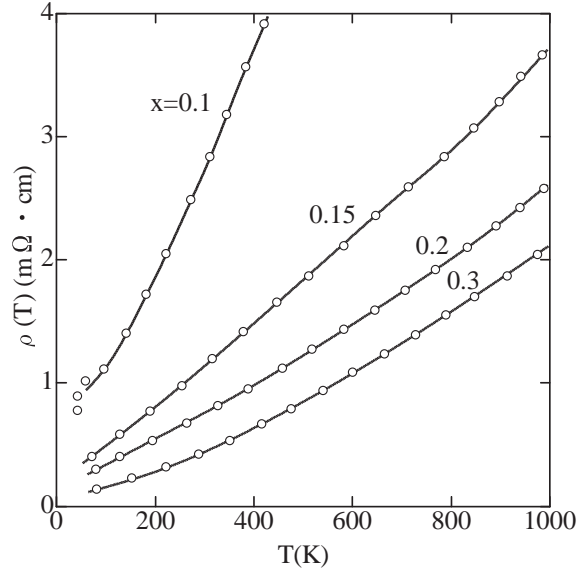}
\caption{ 
$\rho(T)$ vs $T$ of polycrystalline 
       ${\rm La}_{2-x}{\rm Sr}_{x}{\rm Cu}{\rm O}_{4}$.
       Open circles; representative points of the experimental date \cite{AC93}:
       Solid lines; theoretical curves.
}\label{figurename}\end{center}\end{figure}

\begin{figure}[btp]
\begin{center}\leavevmode
\includegraphics[width=\linewidth]{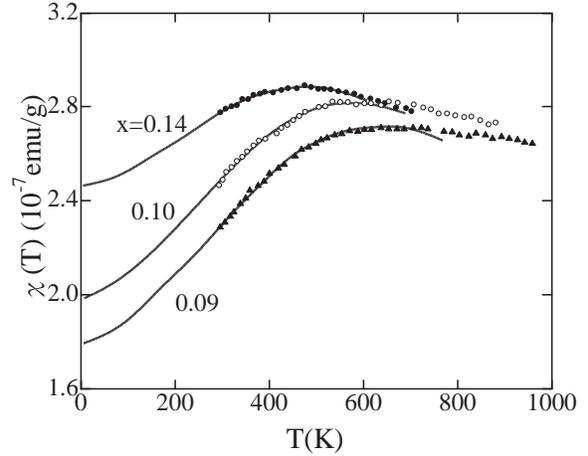}
\caption{ 
$\chi (T)$ vs $T$ of 
        ${\rm La}_{2-x}{\rm Sr}_{x}{\rm Cu}{\rm O}_{4}$. Symbols; experimental data \cite{AC92}: 
        Solid lines; theoretical curves.
}\label{figurename}\end{center}\end{figure}

\end{document}